\documentclass[twocolumn,showpacs,preprintnumbers,amsmath,amssymb]{revtex4-1}

\usepackage{graphicx}
\usepackage{dcolumn}
\usepackage{bm}

\begin{document}

\newtheorem{theorem}{Theorem}
\newtheorem{definition}{Definition}
\newtheorem{lemma}{Lemma}
\newtheorem{proposition}{Proposition}
\newtheorem{remark}{Remark}
\newtheorem{corollary}{Corollary}
\newtheorem{example}{Example}


\title{Fractional diffusion on a fractal grid comb}

\author{Trifce Sandev}
\email{sandev@pks.mpg.de} \affiliation{Max Planck Institute for the Physics of Complex Systems,
N\"{o}thnitzer Strasse 38, 01187 Dresden, Germany} \affiliation{Radiation Safety
Directorate, Partizanski odredi 143, P.O. Box 22, 1020 Skopje,
Macedonia}

\author{Alexander Iomin}
\email{iomin@physics.technion.ac.il} \affiliation{Max Planck Institute for the Physics of
Complex Systems, N\"{o}thnitzer Strasse 38, 01187 Dresden, Germany} \affiliation{Department of Physics, Technion, Haifa 32000, Israel}

\author{Holger Kantz}
\email{kantz@pks.mpg.de} \affiliation{Max Planck Institute for the Physics of Complex Systems,
N\"{o}thnitzer Strasse 38, 01187 Dresden, Germany}

\date{Received 20 October 2014, Published 4 March 2015}

\begin{abstract}
A grid comb model is a generalization of the well known comb
model, and it consists of $N$ backbones. For $N=1$ the system
reduces to the comb model where subdiffusion takes place with the
transport exponent $1/2$. We present an exact analytical
evaluation of the transport exponent of anomalous diffusion for
finite and infinite number of backbones. We show that for an
arbitrarily large but finite number of backbones the transport
exponent does not change. Contrary to that, for an infinite number
of backbones, the transport exponent depends on the fractal
dimension of the backbone structure.
\end{abstract}

\pacs{87.19.L-, 05.40.Fb, 82.40.-g}
\keywords{fractal grid comb, mean square displacement, anomalous
diffusion, fractal dimension}
\maketitle

\section{Introduction}

The comb-like models have been introduced to investigate anomalous
diffusion in low-dimensional percolation clusters
\cite{white,weiss,havlin,arkhincheev}. It means that the mean
square displacement (MSD) has power-law dependence on time
$\left\langle x^{2}(t)\right\rangle\simeq t^{\alpha}$
\cite{metzler report}. An elegant form of equation which describes
the diffusion on a comb-like structure was introduced by
\cite{arkhincheev}
\begin{eqnarray}\label{diffusion like eq on a comb}
\frac{\partial}{\partial t}P(x,y,t)&&
=\mathcal{D}_{x}\delta(y)\frac{\partial^{2}}{\partial x^{2}}P(x,y,t)+\mathcal{D}_{y}\frac{\partial^{2}}{\partial y^{2}}P(x,y,t),\nonumber\\
\end{eqnarray}
where $P(x,y,t)$ is the probability distribution function (PDF),
$\mathcal{D}_{x}\delta(y)$ is the diffusion coefficient in the $x$
direction with physical dimension
$[\mathcal{D}_{x}]=\mathrm{m}^{3}/\mathrm{s}$, and
$\mathcal{D}_{y}$ is the diffusion coefficient in the $y$ direction
with physical dimension
$[\mathcal{D}_{y}]=\mathrm{m}^{2}/\mathrm{s}$. The
$\delta$ function in the diffusion coefficient in the $x$ direction
implies that the diffusion along the $x$ direction occurs only at
$y=0$. Thus, this equation can be used to describe diffusion in
the backbone (at $y=0$) where the teeth play the role of traps.

Nowadays,  comb models have many applications. They have been used
for the understanding of continuous \cite{arkhincheev chaos,iomin
prl2004,da silva} and discrete \cite{cassi} non-Markovian random
walks. There are generalizations of this equation by introducing
time fractional derivatives and integrals in (\ref{diffusion like
eq on a comb}) \cite{mendez,iomin}. Such generalized comb-like
models have been used to describe anomalous diffusion in spiny
dendrites, where the MSD along the $x$ direction has a power-law
dependence on time \cite{mendez,iomin}, or for describing
subdiffusion on a fractal comb \cite{iomin2}, the mechanism of
superdiffusion of ultra-cold atoms in a one dimensional
polarization optical lattice \cite{iomin3} as a phenomenology of
experimental study \cite{sagi}, and to describe diffusion
processes on a backbone structure \cite{lenzi}. Different
generalizations of the comb model have been shown to represent
more realistic models for describing transport properties in
discrete systems, such as porous discrete media \cite{maex},
electronic transport in semiconductors with a discrete
distribution of traps, cancer development with definitely fractal
structure of the spreading front \cite{sokolov,gouyet},
infiltration of diffusing particles from one material to another
\cite{korabel}, description of diffusion of active species in
porous media \cite{AKB}, etc. Furthermore, in \cite{iomin pre2005}
it is shown that in a comb-like model a negative superdiffusion
occurs due to the presence of an inhomogeneous convection flow.

In this paper we consider a generalization of Eq.(\ref{diffusion like eq
  on a comb})
where we allow that diffusion along the $x$ direction may occur on
many backbones, located at $y=l_{j}$, $j=1,2,\dots,N$, $0\leq
l_{1}<l_{2}<\dots<l_{N}$. This means that we have a comb grid where
$N$ can be arbitrarily large, even infinity. The governing
equation for such a structure is given by
{\small{\begin{eqnarray}\label{diffusion like eq on a comb two
delta} \frac{\partial}{\partial t}P(x,y,t)
=&&\mathcal{D}_{x}\sum_{j=1}^{N}w_{j}\delta(y-l_{j})\frac{\partial^{2}}{\partial
x^{2}}P(x,y,t)\nonumber\\&&+\mathcal{D}_{y}\frac{\partial^{2}}{\partial
y^{2}}P(x,y,t),
\end{eqnarray}}}where $w_{j}$ are structural constants such that $\sum_{j=1}^{N}w_{j}=1$.
The initial condition is given by
\begin{equation}
\label{initial condition}
P(x,y,t=0)=\delta(x)\delta(y),
\end{equation}
and the boundary conditions for $P(x,y,t)$ and $\frac{\partial}{\partial
  q}P(x,y,t)$, $q=\{x,y\}$ are set to zero at infinity, $x=\pm\infty$,
$y=\pm\infty$. One can easily verify that for $l_{1}=0$,
$w_{1}=1$, and $w_{2}=w_{3}=\dots=w_{N}=0$ Eq.(\ref{diffusion
like eq on a comb two delta}) becomes (\ref{diffusion like eq on a
comb}). The physical dimensions of $\mathcal{D}_{x}$ and
$\mathcal{D}_{y}$ for a finite number of backbones are the same as those
in Eq.(\ref{diffusion like eq on a comb}). The case of a
fractal structure of backbones will be described by an appropriate
generalization of Eq.(\ref{diffusion like eq on a comb two
delta}). The motivation to introduce such a model is to describe the diffusion of solvents in thin porous films \cite{Shamiryan}. Such
a product structure of backbones times comb is an idealization of
more complex comb-like fractal networks, as they may appear, e.g.,
in certain anisotropic porous media or anisotropic biological
tissue.

The paper is organized as follows. In Sec. II we analyze the
PDF and the MSD in both directions for the force free case.
Anomalous diffusive behavior $\left\langle
x^{2}(t)\right\rangle\simeq t^{1/2}$ appears in the $x$ direction due
to the comb structure of the system. General results for the MSD
in the case of a finite number of backbones $N$ are presented. We also
investigate the effects of an external constant force, applied
along the backbones, on the particle behavior. In Sec. III we
consider an infinite number of backbones. It is shown that an
infinite number of backbones, different from the case of a finite
number of backbones, changes the transport exponent. Deviations
from the standard MSD $\left\langle x^{2}(t)\right\rangle\simeq
t^{1/2}$ were observed recently in combs with ramified teeth as
well, due to teeth with a fractal structure \cite{barkai}. The
summary is given in Sec. IV.

\section{Finite number of backbones. MSD}

We apply a Laplace transform ($\mathcal{L}[f(t)]=\hat{f}(s)$) to Eq.(\ref{diffusion like eq on a comb two delta}), and then a
Fourier transform with respect to the $x$
($\mathcal{F}_{x}[f(x)]=\tilde{f}(\kappa_{x})$) and $y$
($\mathcal{F}_{y}[f(y)]=\bar{f}(\kappa_{y})$) variables. Thus, we
obtain
\begin{eqnarray}\label{diffusion like eq on a comb Fourier-Laplace transform}
\bar{\tilde{\hat{P}}}(\kappa_{x},\kappa_{y},s)=&&\frac{\bar{\tilde{P}}(\kappa_{x},\kappa_{y},t=0)}{s+\mathcal{D}_{y}\kappa_{y}^{2}}\nonumber\\&&-\frac{\sum_{j=1}^{N}w_{j}\tilde{\hat{P}}(\kappa_{x},y=l_{j},s)\exp\left(i\kappa_{y}l_{j}\right)}{s+\mathcal{D}_{y}\kappa_{y}^{2}}\mathcal{D}_{x}\kappa_{x}^{2},\nonumber\\
\end{eqnarray}
where $\bar{\tilde{P}}(\kappa_{x},\kappa_{y},t=0)=1$. From relation (\ref{diffusion like eq on a comb Fourier-Laplace transform}), the inverse Fourier transform with respect to $\kappa_{y}$ yields
\begin{eqnarray}\label{diffusion like eq on a comb Fourier-Laplace transform inverse F ky}
&&\tilde{\hat{P}}(\kappa_{x},y,s)=\frac{\exp\left(-\sqrt{\frac{s}{\mathcal{D}_{y}}}|y|\right)}{2\sqrt{\mathcal{D}_{y}}s^{1/2}}\nonumber\\&&-\frac{\mathcal{D}_{x}\kappa_{x}^{2}\sum_{j=1}^{N}w_{j}\tilde{\hat{P}}(\kappa_{x},y=l_{j},s)\exp\left(-\sqrt{\frac{s}{\mathcal{D}_{y}}}|y-l_{j}|\right)}{2\sqrt{\mathcal{D}_{y}}s^{1/2}}.\nonumber\\
\end{eqnarray}

In the setting of a comb model, the nontrivial and interesting motion is along the backbones, i.e., along the $x$ direction, while the $y$ direction is an auxiliary subspace. Therefore, integrating the motion in the $y$ direction, we analyze the PDF $p_{1}(x,t)=\int_{-\infty}^{\infty}dyP(x,y,t)$. By integration of Eq.(\ref{diffusion like eq on a comb two delta}) with respect to $y$ and performing the Laplace transform with respect to time $t$, and the Fourier transform with respect to $x$, one obtains
\begin{eqnarray}\label{p1 general}
\tilde{\hat{p}}_{1}(\kappa_{x},s)=\frac{1}{s}\left[1-\mathcal{D}_{x}\kappa_{x}^{2}\sum_{j=1}^{N}w_{j}\tilde{\hat{P}}(\kappa_{x},y=l_{j},s)\right].
\end{eqnarray}

From the PDF (\ref{p1 general}) we calculate the MSD along the $x$ direction by the following formula:
\begin{eqnarray}\label{MSD general}
\left\langle x^{2}(t)\right\rangle=\mathcal{L}^{-1}\left.\left[-\frac{\partial^{2}}{\partial\kappa_{x}^{2}}\tilde{\hat{p}}_{1}(\kappa_{x},s)\right]\right|_{\kappa_{x}=0}.
\end{eqnarray}
From relations (\ref{diffusion like eq on a comb Fourier-Laplace transform inverse F ky})-(\ref{MSD general}) for the MSD we derive
\begin{eqnarray}\label{diffusion like eq on a comb MSD exact N}
\left\langle x^{2}(t)\right\rangle&&=\frac{\mathcal{D}_{x}}{\sqrt{\mathcal{D}_{y}}}\mathcal{L}^{-1}\left[s^{-3/2}\sum_{j=1}^{N}w_{j}e^{-\sqrt{\frac{s}{\mathcal{D}_{y}}}|l_{j}|}\right]\nonumber\\&&=\frac{\mathcal{D}_{x}}{\sqrt{\mathcal{D}_{y}}}\sum_{j=1}^{N}w_{j}\left[\frac{2}{\sqrt{\pi}}t^{1/2}e^{-\frac{|l_{j}|^{2}}{4\mathcal{D}_{y}t}}\right.\nonumber\\&&\left.-\frac{|l_{j}|}{\sqrt{\mathcal{D}_{y}}}\mathrm{erfc}\left(\frac{|l_{j}|}{\sqrt{4\mathcal{D}_{y}t}}\right)\right],
\end{eqnarray}
where $\mathrm{erfc}(x)$ is the complementary error function
$\mathrm{erfc}(x)=\frac{2}{\sqrt{\pi}}\int_{x}^{\infty}due^{-u^{2}}$
\cite{erdelyi}.

For $l_{1}=0$ it follows that
\begin{eqnarray}\label{diffusion like eq on a comb MSD exact N l1-0}
\left\langle x^{2}(t)\right\rangle =&&\frac{2w_{1}\mathcal{D}_{x}}{\sqrt{\mathcal{D}_{y}}}\frac{t^{\frac{1}{2}}}{\Gamma\left(\frac{1}{2}\right)}+\frac{\mathcal{D}_{x}}{\sqrt{\mathcal{D}_{y}}}\nonumber\\&&\times\sum_{j=2}^{N}w_{j}\left[\frac{2}{\sqrt{\pi}}t^{1/2}e^{-\frac{|l_{j}|^{2}}{4\mathcal{D}_{y}t}}\right.\nonumber\\&&\left.-\frac{|l_{j}|}{\sqrt{\mathcal{D}_{y}}}\mathrm{erfc}\left(\frac{|l_{j}|}{\sqrt{4\mathcal{D}_{y}t}}\right)\right].
\end{eqnarray}

For the long time scale when
$\frac{|l_{j}|}{\sqrt{\mathcal{D}_{y}t}}\ll1$, $j=2,3,\dots,N$, the
MSD reads
\begin{eqnarray}\label{diffusion like eq on a comb MSD exact l1=0,N t->large}
\left\langle x^{2}(t)\right\rangle=\frac{2\sum_{j=1}^{N}w_{j}\mathcal{D}_{x}}{\sqrt{\mathcal{D}_{y}}}\frac{t^{\frac{1}{2}}}{\Gamma\left(\frac{1}{2}\right)},
\end{eqnarray}
which means that all backbones contribute in the MSD. In contrast
to this, on a short time scale, when
$\frac{|l_{j}|}{\sqrt{\mathcal{D}_{y}t}}\gg1$, $j=2,3,\dots,N$, one
finds that the main contribution in the MSD is due to the first
backbone, i.e.,
\begin{eqnarray}\label{diffusion like eq on a comb MSD exact l1=0,L->infty}
\left\langle x^{2}(t)\right\rangle\simeq\frac{2w_{1}\mathcal{D}_{x}}{\sqrt{\mathcal{D}_{y}}}\frac{t^{\frac{1}{2}}}{\Gamma\left(\frac{1}{2}\right)},
\end{eqnarray}
This result is expected since for short times the particles move
mainly in the first backbone because they had not enough time to
reach the other ones by diffusion in the $y$ direction. This can
be easily verified by considering diffusion along the $y$ direction. We analyze the PDF $p_{2}(y,t)=\int_{-\infty}^{\infty}dxP(x,y,t)$,
for which we find that
\begin{equation}\label{p2 general}
\bar{\hat{p}}_{2}(\kappa_{y},s)=\frac{1}{s+\mathcal{D}_{y}\kappa_{y}^{2}},
\end{equation}
i.e.,
$p_{2}(y,t)=\frac{1}{\sqrt{4\pi\mathcal{D}_{y}t}}\exp\left(-\frac{y^{2}}{4\mathcal{D}_{y}t}\right)$.
For the MSD along the $y$ direction one finds a linear dependence on time $\left\langle y^{2}(t)\right\rangle=2\mathcal{D}_{y}t$, i.e.,
normal diffusion along the $y$ direction. Therefore, the
probability to find the particle at the first backbone is
$p_{2,1}(y,t)=\frac{1}{\sqrt{4\pi\mathcal{D}_{y}t}}$ ($l_{1}=0$), while at the second backbone it is
$p_{2,2}(y,t)=\frac{1}{\sqrt{4\pi\mathcal{D}_{y}t}}
\exp\left(-\frac{l_{2}^{2}}{4\mathcal{D}_{y}t}\right)$,
and so on. Since for the short time scales, $p_{2,1}(y,t)\gg
p_{2,2}(y,t)\gg\dots$, we conclude that the main contribution in the MSD for short times is due to the displacements in the first backbone.

From relation (\ref{diffusion like eq on a comb MSD exact N l1-0}) for $w_{1}=1$, $w_{2}=w_{3}=\dots=w_{N}=0$, and $l_{1}=0$ (which means one backbone) we obtain the MSD for the comblike model
(\ref{diffusion like eq on a comb})
\begin{eqnarray}\label{diffusion eq on a comb MSD}
\left\langle x^{2}(t)\right\rangle=\frac{2\mathcal{D}_{x}}{\sqrt{\mathcal{D}_{y}}}\frac{t^{\frac{1}{2}}}{\Gamma\left(\frac{1}{2}\right)}.
\end{eqnarray}

\begin{figure}
\resizebox{0.5\textwidth}{!}{\includegraphics{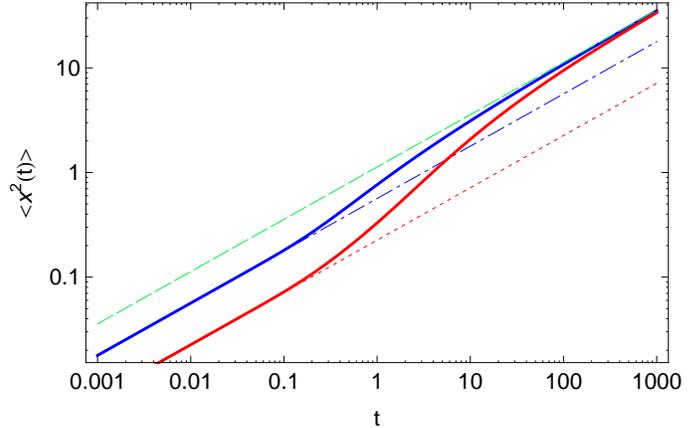}}
\caption {(Color online) Graphical representation of the MSD (\ref{diffusion like
eq on a comb MSD exact N l1-0}) on log-log scale. The blue solid
line (upper solid line) corresponds to the MSD in the case of two
backbones, $l_{1}=0$, $l_{2}=L=1$, and $w_{1}=w_{2}=1/2$. The blue
dot-dashed line describes the asymptotic behavior of the MSD for
short times, given by (\ref{diffusion like eq on a comb MSD exact
l1=0,L->infty}). The red solid line (lower solid line) corresponds
to the MSD in case of five backbones, $l_{j}=(j-1)L$,
$j=1,2,\dots,5$, $L=1$, $w_{j}=1/5$. The red dotted line corresponds
to its asymptotic behavior for short times, given by
(\ref{diffusion like eq on a comb MSD exact l1=0,L->infty}). The
MSDs in both cases have the same asymptotic in the long time limit
given by (\ref{diffusion like eq on a comb MSD exact l1=0,N
t->large}), i.e. $\left\langle
x^{2}(t)\right\rangle=
\frac{2\mathcal{D}_{x}}{\sqrt{\mathcal{D}_{y}}}\frac{t^{1/2}}{\Gamma\left(1/2\right)}$
(green dashed line). Diffusion coefficients are set to 1;
$\mathcal{D}_{x}=\mathcal{D}_{y}=1$.} \label{comb}
\end{figure}

These results are supported by graphical representation in Fig.
\ref{comb} of the MSD in the case of two backbones and five backbones. It is assumed that the first backbone is at $y=0$ and all the other backbones are at distances equal to $L$, $2L$, $3L$, $4L$.

From relations (\ref{diffusion like eq on a comb MSD exact l1=0,N
t->large}) and (\ref{diffusion like eq on a comb MSD exact
l1=0,L->infty}) we conclude that any finite number of backbones
does not change the transport exponent in the short and long time
limit. In the intermediate times there is more complicated
behavior of the MSD given by relation (\ref{diffusion like eq on a
comb MSD exact N l1-0}). The crossover time scales separating the
behavior at short, intermediate, and long times are given by
$t_{short}=\min\{l_{j}^{2},j>1\}/2\mathcal{D}_{y}=l_{2}^{2}/2\mathcal{D}_{y}$
and
$t_{long}=\max\{l_{j}^{2}\}/2\mathcal{D}_{y}=l_{N}^{2}/2\mathcal{D}_{y}$.

In the presence of a constant external force $F$ along the
backbones we arrive at the following Fokker-Planck equation
\begin{eqnarray}\label{FPeq on a comb two delta}
\frac{\partial}{\partial t}P(x,y,t)
=&&\sum_{j=1}^{N}w_{j}\delta(y-l_{j})\left[-\eta F\frac{\partial}{\partial x}+\mathcal{D}_{x}\frac{\partial^{2}}{\partial x^{2}}\right]\nonumber\\&&\times P(x,y,t)+\mathcal{D}_{y}\frac{\partial^{2}}{\partial y^{2}}P(x,y,t),
\end{eqnarray}
where $\eta$ is the mobility. One can compute the first moment as a function of time,
\begin{eqnarray}\label{FPeq on a comb first moment}
\left\langle x(t)\right\rangle_{F}
=\frac{\eta F}{2\sqrt{\mathcal{D}_{y}}}\mathcal{L}^{-1}\left[s^{-3/2}\sum_{j=1}^{N}w_{j}e^{-\sqrt{\frac{s}{\mathcal{D}_{y}}}|l_{j}|}\right],\nonumber\\
\end{eqnarray}
where by comparing it with relation (\ref{diffusion like eq on a comb MSD exact N})
we conclude that the generalized Einstein relation is fulfilled \cite{metzler report},
\begin{eqnarray}\label{Einstein relation}
\left\langle x(t)\right\rangle_{F}=\frac{F}{2k_{B}T}\left\langle x^{2}(t)\right\rangle_{F=0},
\end{eqnarray}
where $\eta k_{B}T=\mathcal{D}_{x}$.

\section{Fractal structure of backbones}

To introduce a fractal structure of the backbones we go back to
Eq.(\ref{diffusion like eq on a comb two delta}) and replace
the summation $\sum_{j=1}^{N}w_{j}\delta(y-l_{j})$ with summation
over a fractal set $\mathcal{S}_{\nu}$, i.e., $\sum_{l_{j}\in
\mathcal{S}_{\nu}}\delta(y-l_{j})$, which means that the backbones
are at positions $y$ which belong to the fractal set
$\mathcal{S}_{\nu}$ with fractal dimension $0<\nu<1$.

A simple toy example, which corresponds to an infinite fractal
set, can be treated as follows. In relation (\ref{diffusion like eq on a comb MSD exact N}) we calculate
$\sum_{j=1}^{N}w_{j}e^{-\sqrt{\frac{s}{\mathcal{D}_{y}}}|l_{j}|}\rightarrow
\sum_{l_{j}\in
\mathcal{S}_{\nu}}e^{-\sqrt{\frac{s}{\mathcal{D}_{y}}}|l_{j}|}$. One should recognize that fractal sets (like a Cantor set) are uncountable. Therefore, the last expression is purely formal and its mathematical realization corresponds to integration to fractal measure $\mu_{\nu}\sim l^{\nu}$ such that $\sum_{l_{j}\in
\mathcal{S}_{\nu}}\delta(l-l_{j})=\frac{1}{\Gamma(\nu)}l^{\nu-1}$
is the fractal density \cite{SKM book,tarasov}, and
$d\mu_{\nu}=\frac{1}{\Gamma(\nu)}l^{\nu-1}dl$. Here we note that $\mathcal{D}_{x}$ is a generalized diffusion coefficient with physical dimension
$[\mathcal{D}_{x}]=\mathrm{m}^{3-\nu}/\mathrm{s}$ that absorbs
the dimension of fractal volume or measure $\mu_{\nu}$. That finally yields the following integration:
\begin{eqnarray}\label{infinite fractal mesh}
\frac{1}{\Gamma(\nu)}\int_{0}^{\infty}dl\,l^{\nu-1}e^{-\sqrt{\frac{s}{\mathcal{D}_{y}}}l}=\left(\frac{\mathcal{D}_{y}}{s}\right)^{\nu/2}.
\end{eqnarray}
For the MSD, we obtain from (\ref{diffusion like eq on a comb MSD exact N})
\begin{equation}\label{MSD infinite fractal mesh}
\left\langle x^{2}(t)\right\rangle=\frac{\mathcal{D}_{x}}{\mathcal{D}_{y}^{\frac{1-\nu}{2}}}\frac{t^{\frac{1+\nu}{2}}}{\Gamma\left(1+\frac{1+\nu}{2}\right)},
\end{equation}
i.e., anomalous diffusive behavior with the transport exponent equal to
$\frac{1}{2}<\frac{1+\nu}{2}<1$. Thus, the fractal set $\mathcal{S}_{\nu}$ of
the infinite number of backbones changes the transport exponent, from $1/2$ to
$\frac{1+\nu}{2}$. For $\nu=1$ the MSD becomes $\left\langle
  x^{2}(t)\right\rangle\simeq t$, which is consistent with
expectations, and for $\nu=0$, we are back to the finite-$N$ case. Indeed, the fractal dimension of any finite number of discrete points is $\nu=0$.

We further consider a random fractal set $\mathcal{S}_{\nu}\in[a,b]$, with finite limits. From relation
(\ref{diffusion like eq on a comb MSD exact N}), in the same way as in (\ref{infinite fractal mesh}), for a finite integration in $[0,L]$, one finds a result in the form of an incomplete $\gamma$
function $\gamma(a,x)=\int_{0}^{x}dt\,t^{a-1}e^{-t}$
\cite{erdelyi},
\begin{equation}\label{finite integration}
\frac{1}{\Gamma(\nu)}\int_{0}^{L}dl\,l^{\nu-1}e^{-\sqrt{\frac{s}{\mathcal{D}_{y}}}l}=\left(\frac{\mathcal{D}_{y}}{s}\right)^{\nu/2}\frac{\gamma(\nu,L)}{\Gamma(\nu)}.
\end{equation}
Thus, the MSD becomes
\begin{equation}\label{MSD finite fractal mesh}
\left\langle x^{2}(t)\right\rangle=\frac{\mathcal{D}_{x}}{\mathcal{D}_{y}^{\frac{1-\nu}{2}}}\frac{\gamma(\nu,L)}{\Gamma(\nu)}\frac{t^{\frac{1+\nu}{2}}}{\Gamma\left(1+\frac{1+\nu}{2}\right)}.
\end{equation}
Again, for $\nu=1$ the normal diffusive behavior along the
$x$ direction appears, i.e., $\left\langle
x^{2}(t)\right\rangle\simeq t$.

Here we note that the result for the MSD (\ref{MSD infinite
fractal mesh}) can be obtained in the framework of fractional
integration as well. By integration of Eq.(\ref{diffusion
like eq on a comb two delta}) over $y$ and using the summation on the fractal set as above in this section, for the PDF $p_{1}(x,t)$ one obtains
\begin{equation}\label{p1 fractal}
\frac{\partial}{\partial t}p_{1}(x,t)=\mathcal{D}_{x}\sum_{l_{j}\in \mathcal{S}_{\nu}}\frac{\partial^{2}}{\partial x^{2}}p(x,y=l_{j},t).
\end{equation}
The Laplace transform to (\ref{p1 fractal}) yields
\begin{equation}\label{p1 fractal Laplace}
s\hat{p}_{1}(x,s)-p_{1}(x,t=0)=\mathcal{D}_{x}\sum_{l_{j}\in \mathcal{S}_{\nu}}\frac{\partial^{2}}{\partial x^{2}}\hat{p}(x,y=l_{j},s).
\end{equation}
By representing the solution $p(x,y,s)$ in the following way:
$\hat{p}(x,y,s)=\hat{g}(x,s)e^{-\sqrt{\frac{s}{\mathcal{D}_{y}}}|y|}$,
i.e.,
$\hat{p}(x,y=l_{j},s)=\hat{g}(x,s)e^{-\sqrt{\frac{s}{\mathcal{D}_{y}}}|l_{j}|}$,
for the $\hat{p}_{1}(x,s)$ we find
\begin{equation}\label{p1 new}
\hat{p}_{1}(x,s)=\int_{-\infty}^{\infty}dyp(x,y,s)=2\hat{g}(x,s)\sqrt{\frac{\mathcal{D}_{y}}{s}}.
\end{equation}
From the other side, by using the previous approach of summation, we have
\begin{eqnarray}\label{summation p lj}
\sum_{l_{j}\in\mathcal{S}_{\nu}}\hat{p}(x,y=l_{j},s)&&=\hat{g}(x,s)\frac{1}{\Gamma(\nu)}\int_{0}^{\infty}dll^{\nu-1}e^{\sqrt{\frac{s}{\mathcal{D}_{y}}}l}\nonumber\\&&=\hat{g}(x,s)\left(\frac{\mathcal{D}_{y}}{s}\right)^{\nu/2}\nonumber\\&&=\frac{1}{2\mathcal{D}_{y}^{\frac{1-\nu}{2}}}s^{\frac{1-\nu}{2}}\hat{p}_{1}(x,s).
\end{eqnarray}
By substituting relation (\ref{summation p lj}) in (\ref{p1 fractal Laplace}), we obtain
\begin{eqnarray}\label{p1 fractal Laplace final}
&&s^{\frac{1+\nu}{2}}\hat{p}_{1}(x,s)-s^{\frac{1+\nu}{2}-1}p_{1}(x,t=0)\nonumber\\&&=\frac{\mathcal{D}_{x}}{2\mathcal{D}_{y}^{\frac{1-\nu}{2}}}\frac{\partial^{2}}{\partial x^{2}}\hat{p}_{1}(x,s).
\end{eqnarray}
From this, the inverse Laplace transform yields the following time fractional diffusion equation:
\begin{equation}\label{p1 fractal Laplace final}
\frac{\partial^{\frac{1+\nu}{2}}}{\partial t^{\frac{1+\nu}{2}}}p_{1}(x,t)=\frac{\mathcal{D}_{x}}{2\mathcal{D}_{y}^{\frac{1-\nu}{2}}}\frac{\partial^{2}}{\partial x^{2}}p_{1}(x,t),
\end{equation}
where $\frac{\partial^{\frac{1+\nu}{2}}}{\partial
t^{\frac{1+\nu}{2}}}$ is the Caputo time fractional derivative of
order $\frac{1}{2}<\frac{1+\nu}{2}<1$ \cite{Caputo,caputo
derivative}. From here we easily obtain the MSD $\left\langle
x^{2}(t)\right\rangle=\int_{-\infty}^{\infty}dxx^{2}p_{1}(x,t)$
that is of form (\ref{MSD infinite fractal mesh}). The solution
for the PDF $p_{1}(x,t)$ can be represented in terms of the Fox
$H$ function $H_{p,q}^{m,n}(z)$ \cite{saxena book,sandev jpa2011},
\begin{equation}\label{p1 fractal structure}
p_{1}(x,t)=\frac{1}{2|x|}H_{1,1}^{1,0}\left[\left.\frac{|x|}{\sqrt {\mathcal{D}_{\nu}t^{(1+\nu)/2}}}\right|\begin{array}{c l}
    (1,(1+\nu)/4)\\
    (1,1)
  \end{array}\right],
\end{equation}
where
$\mathcal{D}_{\nu}=\mathcal{D}_{x}/2\mathcal{D}_{y}^{\frac{1-\nu}{2}}$
is the generalized diffusion coefficient with physical dimension
$[\mathcal{D}_{\nu}]=\mathrm{m}^{2}/\mathrm{s}^{(1+\nu)/2}$.
Therefore, as shown, the infinite number of backbones changes the
transport exponent.

The asymptotic behavior of $p_{1}(x,t)$ (\ref{p1 fractal
structure}) for $\frac{|x|}{\sqrt
{\mathcal{D}_{\nu}t^{(1+\nu)/2}}}\gg1$ is of the form \cite{metzler
report,sandev jpa2011}:
\begin{widetext}
\begin{eqnarray}\label{asymptotic p1 fractal}
p_{1}(x,t)\simeq&&\frac{1}{\sqrt{2(3-\nu)\pi}}\left(\frac{1+\nu}{4}\right)^{\frac{\nu-1}{3-\nu}}|x|^{\frac{\nu-1}{3-\nu}}\left(\mathcal{D}_{\nu}t^{\frac{1+\nu}{2}}\right)^{-\frac{1}{3-\nu}}\nonumber\\&&\times\exp\left[-\frac{3-\nu}{4}\left(\frac{1+\nu}{4}\right)^{\frac{1+\nu}{3-\nu}}|x|^{\frac{4}{3-\nu}}\left(\mathcal{D}_{\nu}t^{\frac{1+\nu}{2}}\right)^{-\frac{2}{3-\nu}}\right],
\end{eqnarray}
\end{widetext}
i.e., it has non-Gaussian behavior. For $\nu=1$ it turns to
Gaussian behavior as it is expected and as it was shown by
analysis of the MSD.

Additionally to the MSD we calculate the $q$th moment
$\left\langle
|x|^{q}\right\rangle=2\int_{0}^{\infty}dx\,x^{q}p_{1}(x,t)$, for
which one finds \cite{metzler report,sandev jpa2011}
\begin{eqnarray}\label{q moment}
\left\langle |x|^{q}\right\rangle=\left(\mathcal{D}_{\nu}t^{\frac{1+\nu}{2}}\right)^{q/2}\frac{\Gamma(1+q)}{\Gamma\left(1+\frac{1+\nu}{2}\frac{q}{2}\right)}.
\end{eqnarray}
Thus for the fourth moment it follows that
\begin{eqnarray}\label{4th moment}
\left\langle |x|^{4}\right\rangle=24\mathcal{D}_{\nu}^{2}\frac{t^{1+\nu}}{\Gamma(2+\nu)}=6\frac{\mathcal{D}_{x}^{2}}{\mathcal{D}_{y}^{1-\nu}}\frac{t^{1+\nu}}{\Gamma(2+\nu)}.
\end{eqnarray}
The calculation of the fourth moment is useful to discriminate
subdiffusive processes with identical MSDs, e.g., subdiffusion due
to different fractal structures or different mechanisms
\cite{spanner} (see also \cite{metzler pccp}). For the even
moments we obtain
\begin{eqnarray}\label{even moments}
\left\langle |x|^{2n}\right\rangle=(2n)!\frac{\mathcal{D}_{\nu}^{n}t^{\frac{(1+\nu)n}{2}}}{\Gamma\left(1+\frac{(1+\nu)n}{2}\right)},
\end{eqnarray}
from which we can find the following interesting relation:
\begin{eqnarray}\label{even moments sum}
\sum_{n=0}^{\infty}\frac{\left\langle |x|^{2n}\right\rangle}{(2n)!}=\sum_{n=0}^{\infty}\frac{\mathcal{D}_{\nu}^{n}t^{\frac{(1+\nu)n}{2}}}{\Gamma\left(1+\frac{(1+\nu)n}{2}\right)}=E_{(1+\nu)/2}\left(\mathcal{D}_{\nu}t^{\frac{1+\nu}{2}}\right),\nonumber\\
\end{eqnarray}
where $E_{\alpha}(z)=\sum_{n=0}^{\infty}\frac{z^{n}}{\Gamma(\alpha
n+1)}$ is the one parameter Mittag-Leffler function \cite{saxena
book}.

\section{Weierstrass function and fractional Riesz
derivative}

Finally, we show how the fractal structure $\mathcal{S}_{\nu}$
relates to the fractional Riesz derivative \cite{SKM book}. Let us consider the fractal structure of backbones in Eq.(\ref{diffusion like eq on a comb two delta}) separately. In the
Fourier-Fourier $\left(\kappa_{x},\kappa_{y}\right)$ space it
reads
\begin{eqnarray}\label{eq in FF space}
&&-\mathcal{D}_{x}\kappa_{x}^{2}\sum_{j=1}^{\infty}w_{j}e^{i\kappa_{y}l_{j}}\bar{\tilde{P}}(\kappa_{x},y=l_{j},t)\nonumber\\&&=-\mathcal{D}_{x}\kappa_{x}^{2}\sum_{j=1}^{\infty}w_{j}e^{i\kappa_{y}l_{j}}\frac{1}{2\pi}\int_{-\infty}^{\infty}d\kappa_{y}{'}\bar{\tilde{P}}(\kappa_{x},\kappa_{y}{'},t)e^{-i\kappa_{y}{'}l_{j}}\nonumber\\&&=-\mathcal{D}_{x}\kappa_{x}^{2}\frac{1}{2\pi}\int_{-\infty}^{\infty}d\kappa_{y}{'}\Psi\left(\kappa_{y}-\kappa_{y}{'}\right)\bar{\tilde{P}}(\kappa_{x},\kappa_{y}{'},t),
\end{eqnarray}
where $\Psi\left(\kappa_{y}-\kappa_{y}{'}\right)$ is the
Weierstrass function \cite{berry}. It can be obtained by the
following procedure \cite{shlesinger}: Let us use
$w_{j}=\frac{l-b}{b}\left(\frac{b}{l}\right)^{j}$, where $l,b>1$,
$l-b\ll b$. Thus
\begin{equation}\label{w_j for Weierstrass}
\sum_{j=1}^{\infty}w_{j}=\frac{l-b}{l}\sum_{j=0}^{\infty}\left(\frac{b}{l}\right)^{j}=1.
\end{equation}
Now $l$ and $b$ are dimensionless scale parameters. Therefore
\begin{equation}\label{Weierstrass}
\Psi(z)=\frac{l-b}{b}\sum_{j=1}^{\infty}\left(\frac{b}{l}\right)^{j}\exp\left(i\frac{z}{l^{j}}\right),
\end{equation}
where $l_{j}=L/l^{j}$, and
$z=\left(\kappa_{y}-\kappa_{y}{'}\right)L$, and for convenience,
we choose $l_{1}=L$. From here one obtains
\begin{equation}\label{Weierstrass2}
\Psi(z/l)=\frac{l}{b}\Psi(z)-\frac{l-b}{b}\exp\left(i\frac{z}{l}\right).
\end{equation}
Neglecting the last term since $l-b\ll b$, therefore the scaling
\begin{equation}\label{Weierstrass3}
\Psi(z/l)\simeq\frac{l}{b}\Psi(z),
\end{equation}
means that $\Psi(z)\sim\frac{1}{z^{1+\nu}}$, where
$\nu=\ln\frac{1}{b}/\ln{l}$ is the fractal dimension. Thus, for
relation (\ref{eq in FF space}) we have
\begin{eqnarray}\label{eq in FF space2}
-\mathcal{D}_{x}\kappa_{x}^{2}\frac{L^{-1-\nu}}{2\pi}\int_{-\infty}^{\infty}d\kappa_{y}{'}|\kappa_{y}-\kappa_{y}{'}|^{-1-\nu}\bar{\tilde{P}}(\kappa_{x},\kappa_{y}{'},t).\nonumber\\
\end{eqnarray}
This integration is the Riesz fractional derivative \cite{SKM
book}.

\section{Summary}

In this paper we introduce a diffusion equation for a comb structure where the
displacements in the $x$ direction are possible along many backbones, even an
infinite number of backbones, and we call this system by grid comb. We analyze
the MSD and we show that by adding a finite number of backbones, the transport
exponent in the long time limit does not change. Differently from that, an
infinite number of backbones changes the transport exponent. Considering a
fractal structure of backbones with fractal dimension $\nu$ we obtained the
dependence of the transport exponent on $\nu$. We stress that the performed
analysis is exact--more precisely, that
 the evaluation of the contribution of the fractal structure
 $\mathcal{S}_{\nu}$ to anomalous diffusion is exact.
 Note that the first attempt to take into account a fractal
structure of traps was performed in \cite{iomin2} in the framework
of a coarse graining procedure of the Fokker-Planck equation that
leads to the fractional differentiation in the real space. In
contrast to that, in the present analysis we are able to perform
an exact analysis for the fractal structure $\mathcal{S}_{\nu}$.
This also relates to exact fractional differentiation in the
reciprocal Fourier space.

In conclusion, it should be admitted that a comb model is a toy
model that can be solved exactly and establishes a relation
between geometry and the transport exponent. As is recently found
it also corresponds to the real physical realization in
experiments on calcium transport in spiny dendrites (see
\cite{mendez,iomin} and references therein). The grid-comb model,
suggested here as the generalization of the comb model,
establishes an exact relation between a complicated fractal
geometry and the transport exponent as well. Another strong
motivation of the model, also related to the result, is that in
the framework of this model it is possible to infer an exactly
fractional derivative related to fractal geometry. All these
points are important for the understanding of anomalous transport in
heterogeneous material, in particular to describe diffusion of
solvents in thin porous films \cite{Shamiryan}, or in another
two-dimensional material like graphene \cite{ruzicka}.

\section*{Acknowledgment}

A.I. would like to thank the Max-Planck Institute for the Physics of
Complex Systems in Dresden, Germany for financial support and
hospitality, as well as the support by the Israel Science
Foundation (ISF-1028).



\begin{thebibliography} {}

\bibitem{white}
S.R. White and M. Barma, J. Phys. A: Math. Gen. {\bf17}, 2995 (1984).

\bibitem{weiss}
G.H. Weiss and S. Havlin, Physica A {\bf134}, 474 (1986).

\bibitem{havlin}
O. Matan, S. Havlin, and D. Staufler, J. Phys. A: Math. Gen. {\bf22}, 2867 (1989).

\bibitem{arkhincheev}
V.E. Arkhincheev and E.M. Baskin, Sov. Phys. JETP {\bf73}, 161 (1991).

\bibitem{metzler report}
R. Metzler and J. Klafter, Phys. Rep. {\bf339}, 1 (2000);
J. Phys. A: Math. Gen. {\bf37}, R161 (2004).

\bibitem{arkhincheev chaos}
V.E. Arkhincheev, Chaos, {\bf17}, 043102 (2007).

\bibitem{iomin prl2004}
E. Baskin and A. Iomin, Phys. Rev. Lett. {\bf93}, 120603 (2004).

\bibitem{da silva}
L.R. da Silva, A.A. Tateishi, M.K. Lenzi, E.K. Lenzi, and P.C. da Silva, Braz. J. Phys. {\bf39}, 483 (2009).

\bibitem{cassi}
D. Cassi and S. Regina, Phys. Rev. Lett. {\bf76}, 2914
(1996); G. Baldi, R. Burioni, and D. Cassi, Phys. Rev. E
{\bf70}, 031111 (2004).

\bibitem{mendez}
V. Mendez and A. Iomin, Chaos Solitons Fractals {\bf53}, 46 (2013).

\bibitem{iomin}
A. Iomin and V. Mendez, Phys. Rev. E {\bf88}, 012706 (2013).

\bibitem{iomin2}
A. Iomin, Phys. Rev. E {\bf83}, 052106 (2011).

\bibitem{iomin3}
A. Iomin, Phys. Rev. E {\bf86}, 032101 (2012).

\bibitem{sagi}
Y. Sagi, M. Brook, I. Almog, and N. Davidson, Phys. Rev. Lett. {\bf108}, 093002 (2012).

\bibitem{lenzi}
E.K. Lenzi, L.R. da Silva, A.A. Tateishi, M.K. Lenzi, and H.V.
Ribeiro, Phys. Rev. E {\bf87}, 012121 (2013).

\bibitem{maex}
K. Maex, M.R. Baklanov, D. Shamiryan, F. Lacopi, S.H. Brongersma,
and Z.S. Yanovitskaya, J. Appl. Phys. {\bf93}, 8793 (2003).

\bibitem{sokolov}
I.M. Sokolov, in {\it Encyclopedia of Complexity and Systems
Science}, edited by R.A. Mayers (Springer-Verlag, New York, 2009), p. 309.

\bibitem{gouyet}
J.-F. Gouyet, {\it Physics and Fractal Structures} (Masson, Paris, 1996).

\bibitem{korabel}
N. Korabel and E. Barkai, Phys. Rev. Lett. {\bf104}, 170603 (2010).

\bibitem{AKB}
V.E. Arkhincheev, E. Kunnen, and M.R. Baklanov,
Microelectron. Eng. {\bf88}, 694 (2011).

\bibitem{iomin pre2005}
A. Iomin and E. Baskin, Phys. Rev. E {\bf71}, 061101 (2005).

\bibitem{Shamiryan}
D. Shamiryan, M.R. Baklanov, P. Lyons, S. Beckx, W. Boullart, and
K. Maex, Colloids Surf., A \textbf{300}, 111 (2007).

\bibitem{barkai}
A. Rebenshtok and E. Barkai, Phys. Rev. E {\bf88}, 052126 (2013).

\bibitem{erdelyi}
A. Erdelyi, W. Magnus, F. Oberhettinger and F.G. Tricomi, {\it
Higher Transcedential Functions}, (McGraw-Hill, New York,
1955), Vol. 3.

\bibitem{SKM book}
S.G. Samko, A.A. Kilbas, and O.I. Marichev, {\it Fractional
Integrals and Derivatives: Theory and Applications} (Taylor and
Francis, London, 1993).

\bibitem{tarasov}
V.E. Tarasov, Chaos {\bf14}, 123 (2004).

\bibitem{Caputo}
M. Caputo, {\it Elasticita e Dissipazione} (Zanichelli,
Bologna) 1969.

\bibitem{caputo derivative}
The Caputo fractional derivative of order $0<\mu<1$ is defined by
$\frac{\partial^{\mu}}{\partial
t^{\mu}}f(t)=\frac{1}{\Gamma(1-\mu)}\int_{0}^{t}\mathrm{d}\tau(t-\tau)^{-\mu}
\frac{d}{d\tau}f(\tau)$
\cite{Caputo}. Its Laplace transform is given by
$\mathcal{L}\left[\frac{\partial^{\mu}}{\partial
t^{\mu}}f(t)\right](s)=s^{\mu}\mathcal{L}\left[f(t)\right](s)-s^{\mu-1}f(0+)$.

\bibitem{saxena book}
A.M. Mathai, R.K. Saxena and H.J. Haubold, {\it The $H$-function:
Theory and Applications} (Springer, New York, 2010).

\bibitem{sandev jpa2011}
T. Sandev, R. Metzler, and Z. Tomovski, J. Phys. A: Math. Theor. {\bf44}, 255203 (2011).

\bibitem{spanner}
M. Spanner, F. H\"{o}fling, G. Schr\"{o}der-Turk, K. Mecke, and T.
Franosch, J. Phys.: Condens. Matter {\bf23}, 234120 (2011).

\bibitem{metzler pccp}
R. Metzler, J.-H. Jeon, A.G. Cherstvy, and E. Barkai, Phys.
Chem. Chem. Phys. {\bf16}, 24128 (2014).

\bibitem{berry}
M.V. Berry and Z.V. Lewis, Proc. R. Soc. London Ser. A {\bf370}, 459 (1980).

\bibitem{shlesinger}
M.F. Shlesinger, J. Stat. Phys. {\bf10}, 421 (1974).

\bibitem{ruzicka}
B.A. Ruzicka, S. Wang, L.K. Werake, B. Weintrub,  K.P. Loh, and H. Zhao, Phys. Rev. B {\bf82}, 195414 (2010).


\end{thebibliography}
\end{document}